\documentclass[twocolumn,preprintnumbers,amsmath,amssymb,pra]{revtex4}
\usepackage{epstopdf}
\usepackage[colorlinks,linkcolor=blue,anchorcolor=blue,citecolor=blue]{hyperref}
\usepackage{txfonts}
\usepackage{subfigure}
\usepackage{graphicx,hyperref}
\usepackage{dcolumn}
\usepackage{bm}
\usepackage{extarrows}
\usepackage{setspace}
\usepackage{float}
 \allowdisplaybreaks
\usepackage{multirow} 
\usepackage{booktabs} 

 \allowdisplaybreaks

\begin{document}
\bibliographystyle{plainnat}
\title{Three-dimensional Isotropic Droplets in Rydberg-dressed Bose Gases}
\author{Hao Zhu$^{1}$}
\author{Yong-Yao Li$^{2}$}
\author{Wen-Kai Bai$^{3}$}
\author{Yan-Mei Yu$^{1}$}
\author{Lin Zhuang$^{4}$}
\author{Wu-Ming Liu$^{1,5,6}$}
\altaffiliation{Corresponding author: wliu@iphy.ac.cn}
\affiliation{$^{1}$Beijing National Laboratory for Condensed Matter Physics, Institute of Physics, Chinese Academy of Sciences, Beijing 100190, China}
\affiliation{$^{2}$Guangdong-Hong Kong-Macao Joint Laboratory for Intelligent Micro-Nano Optoelectronic Technology, School of Physics and Optoelectronic Engineering, Foshan University, Foshan 528225, China}
\affiliation{$^{3}$Institute of Modern Physics, Northwest University, Xi'an 710127, China}
\affiliation{$^{4}$School of Physics, Sun Yat-Sen University, Guangzhou 510257, China}
\affiliation{$^{5}$School of Physics, Sun Yat-Sen University, Guangzhou 510257, China}
\affiliation{$^{6}$Songshan Lake Materials Laboratory, Dongguan, Guangdong 523808, China}
\date{\today}
\begin{abstract}
We predict a scheme for the creation of isotropic three-dimensional droplets in Rydbeg-dressed Bose gases, which contain both repulsive contact interactions and attractive van der Waals interactions causing the quantum fluctuation effect non-negligible. We present detailed beyond mean-field calculations with Lee-Huang-Yang correction and demonstrate the existence of isotropic droplets under realistic experimental conditions. Stable droplets possess flat-top density distribution, and their chemical potentials decrease with the particle number expansion towarding a critical value. We distinguish droplets from bright solitons through peak density, width of condensate and quantum depletion calculations. We summarize a phase diagram of realizing droplets, and subsequently highlight the stability of droplets by real time evolution as well as collisions. Our work provides a novel platform for investigating excitation spectrum and superfluid nature of droplets.
\end{abstract}
\maketitle
\section{Introduction}
Three-dimensional (3D) droplets have attracted much attention in ultracold Bose gases where the mean-field-driven collapse is prohibited by a repulsive interaction arising from quantum fluctuations \cite{PhysRevLett.121.173403,PhysRevLett.115.155302,PhysRevLett.126.230404}. Since the mean-field (MF) theory cannot explain the stability of droplets against collapse \cite{Bottcher_2021}, the beyond mean-field (BMF) theory with the quantum fluctuation induced Lee-Huang-Yang (LHY) correction is employed to explain the lab-observational droplets \cite{PhysRevLett.115.155302,PhysRevLett.122.193902,PhysRevLett.117.100401}. Currently, droplets in Bose-Einstein condensates (BECs) can be divided into two categories: (a) single component magnetic gases, in which the MF effects are given by the dipolar and contact interactions, rendering droplets to exhibit an elongated filament shape \cite{PhysRevLett.127.155301,PhysRevLett.119.255302,PhysRevA.98.023618}; (b) binary mixtures of two spin components or heteronuclear atoms, with almost complete cancellation of the MF effects, leaving a small residual attraction compensated by the repulsive LHY energy \cite{PhysRevA.105.033319,PhysRevA.105.033321,PhysRevA.105.063328}. Recent work relating to droplets, i.e., soliton-droplet transitions \cite{PhysRevResearch.3.L012027,PhysRevLett.120.135301,Li_2017}, rotating droplets \cite{PhysRevLett.123.160405,PhysRevLett.126.244101,PhysRevLett.122.193902} and collective excitations \cite{PhysRevA.101.051601,PhysRevA.102.053303,PhysRevA.86.063609} suggests remarkable application potentials. Whereas, the inelastic collision induced self-evaporation in droplets appeals some other platforms where the droplets can live long timescales to be manipulated \cite{PhysRevLett.128.083401,PhysRevLett.124.143401,PhysRevA.104.043301}.
\par
Resonantly excited Rydberg gases enable researchers to systematically study strong interactions in many-body systems, while its short lifetime hinders the experimental manipulations \cite{PhysRevLett.107.153001,PhysRevLett.104.013001,RevModPhys.82.2313}. This shortcoming can be circumvented by weakly dressing the atomic ground states with a small fraction of the Rydberg states that improves the overall lifetime of the system \cite{Dunning_2016}. More specially, the coupling between the ground state and the $nS$ Rydberg exciting state is conceived to introduce repulsive van der Waals (vdW) interactions with the effective coupling $\tilde{C}_6>0$ and red detunings, which can lead to supersolid states \cite{PhysRevLett.104.195302,PhysRevLett.121.030404,PhysRevResearch.2.023290} or superglass phases \cite{PhysRevLett.116.135303}. In addition, with effective coupling $\tilde{C}_6<0$ and blue detuning, isotropic-nonlocal-attractive vdW interactions are believed to dominate \cite{Balewski_2014,Weber_2017,Wu_2021}. With the combined effects of contact interactions and attractive vdW interactions, anisotropic bright solitons can be excited \cite{PhysRevLett.106.170401}. However, we will prove that the consideration of LHY correction could generate isotropic three-dimensional droplets in the Rydberg-dressed Bose gases.
\par
In this Letter, we utilize beyond mean-field theory to predict three-dimensional isotropic droplets in the $^{87}{Rb}$ Rydberg-dressed gases with attractive vdW interactions. In the homogeneous model, analytical formulas for the LHY correction is derived. By tuning the vdW interaction strength and the particle number, it is possible to realize stable droplets where the MF energy is negative while the LHY energy is positive. We develop beyond Gross-Pitaevskii equation (BGPE) for single-component Bose gases and a number of relevant parameters are calculated to characterize the droplets. When compared to bright solitons, droplets exhibit totally different phenomenons with respect to the particle number booming. After summarizing the phase diagram of sharp or flat-top droplets, we simulate the dynamic evolution and collide dynamic of droplets which indicates a long lifetime to be manipulated in experiments.
\begin{figure}[htbp]
\includegraphics[width=8.6cm]{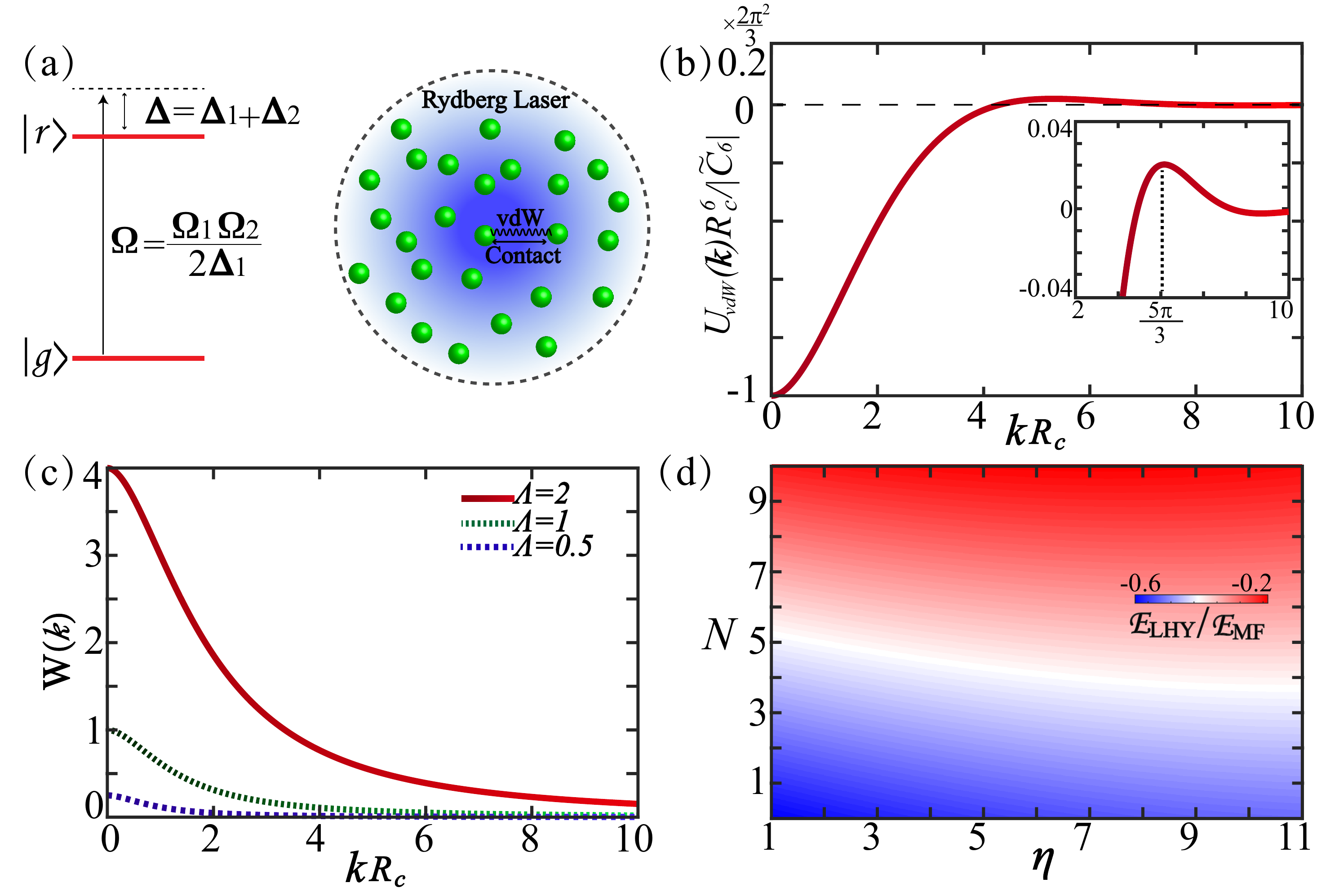}
\caption{(a) Rydberg-dressing states can be realized by imposing BECs into the Rydberg laser. Atoms in BECs interact through MF interactions, including repulsive contact interactions and vdW attractive interactions. The level scheme of the considered Rydberg-dressed approach contain the ground state $|g\rangle$, the Rybderg state $|r\rangle$, and the frequency of the Rydberg laser $\Omega$ with blue detuning $\Delta$. (b) Fourier transformation of the attractive vdW interaction $U_{vdW}({k})$. The minimum of the interaction is located at $kR_c=0$, and the maximum of the interaction is located at $kR_c=\frac{5\pi}{3}$, where $R_c$ represent Rydberg radiuses (explained more clearly from the enlarged view of the $U_{vdW}({k})$). (c) Functions of $W(k)$ vary with momentum ${k}$, which govern the ground-state-energy correction with different $\Lambda$. (d) The LHY energy relative to the mean-field energy, where $\eta=\gamma/\alpha$ equals to the ratio of vdW interaction as well as contact interaction and $N$ is the total particle number.}
\label{fig1}
\end{figure}
\section{Lee-Huang-Yang Corrections in the Rydberg-dressed System}
Our approach bases on optical dressing of ground-state $|g\rangle$ atoms onto highly excited Rydberg states $|r\rangle$ with Rydberg lasers, as illustrated by the simplified two-level model in Fig. 1(a). The blue detuning $\Delta\approx\Delta_1+\Delta_2$ and Rabi frequency $\Omega=\frac{\Omega_1+\Omega_2}{2\Delta_1}$, where $\Omega_{1,2}$ and $\Delta_{1,2}$ are Rabi frequencies and detuning in the two-photon dressing scheme. There are two kinds of MF interactions emerging in this system, i.e., repulsive contact interactions and attractive vdW interactions, which stabilize the condensate from collapse. Inspired by the recent research on the droplets induced by the balance between the MF energy and the LHY energy \cite{PhysRevLett.115.155302,PhysRevLett.117.100401,PhysRevLett.127.203402}, we are willing to produce 3D droplets instead of bright solitons when considering the quantum fluctuation in the Rydberg-dressed system.
\par
We consider $N$ Rydberg-dressed atoms that interact through both $s$-wave contact and vdW long-range interactions. The time-independent ground-state wave function $\Psi_0$ satisfies the equation $\mu \Psi_0=\mathcal{L}_{GP}\Psi_0$, where $\mu$ is the chemical potential and $\mathcal{L}_{GP}=-\frac{\hbar^2\nabla^2}{2M}+\Phi(\bm{r})$. The effective potential $\Phi(\bm{r})=g\delta(\bm{r}-\bm{r}')+\int d^3\bm{r}'U_{vdw}(\bm{r}-\bm{r}')|\Psi(\bm{r}')|^2$ describes the two-body interactions where the coupling constant $g=\frac{4\pi\hbar^2a_s}{M}$ being related to the $s$-wave scattering length $a_s$, the soft-core potential $U_{vdW}(\bm{r}-\bm{r}')=\frac{\tilde{C}_6}{R_c^6+|\bm{r}-\bm{r}'|^6}$ with the effective coupling constant $\tilde{C}_6<0$ and $R_c$ representing the blockade radius which depends on the details of the Rydberg dressing \cite{Balewski_2014,Weber_2017,Wu_2021}. The Fourier transformation of the soft-core potential reads ${U}_{vdW}({k})=U_0f({k})$, where $U_0=\frac{2\pi^2|\tilde{C}_6|}{3R_c^6}$ identifies the strength and $f({k})$ has an analytical form, $f({k})=-e^{-kR_c/2}[e^{-kR_c/2}-2sin(\pi/6-\sqrt{3}kR_c/2)]/kR_c$, which characterizes the momentum dependence of the interaction (see Fig. 1(b)). The minimum value of $U_{vdW}({k})$ locates at $kR_c=0$, while the maximum value appears at $kR_c\simeq5\pi/3$. Consequently, the maximum value of the MF interaction $\Phi(k)$ is $\Phi_{max}(k)=g+U_{vdW}(\frac{5\pi}{3R_c})$ and the minimum value equals to $\Phi_{min}(k)=g+U_{vdW}(0)$.
\par
The ground-state energy correction $\Delta E$ reads: \par
\begin{center}
\begin{equation}
\begin{split}
\Delta\! E\!=\!\frac{1}{2}\!V\!\int\! \frac{d^3k}{(2\pi)^3}\!\left[\!\epsilon_k\!-\!\frac{\hbar^2{k}^2}{2M}\!-\!n\Phi({k})\!+\!\frac{Mn^2}{\hbar^2}\!\frac{{\Phi}^2({k})}{k^2}\!\right]\!,
\end{split}\tag{1}
\end{equation}
\end{center}
\par
\par\noindent where $\epsilon_k$ stands for the Bogoliubov spectrum, ${k}$ represents the wave vector, $n$ is the particle density and $V$ is the volume of the condensate. Supposing $\Phi({k})=\Lambda>0$, and $\Delta E$ can be simplified as $\Delta E=\frac{1}{2}V\int_0^{\infty}W(k)dk$ where $W(k)=k^2\sqrt{\frac{1}{4}k^4+\Lambda k^2}-\frac{1}{2}k^4-\Lambda k^2+\Lambda^2$. With different $\Lambda$ values, $W(k)$ varies with the momentum $k$, as shown in Fig. 1(c). Obviously, $\Delta E$ increases linearly with $\Lambda$ and we can easily obtain the extreme value $\Delta E_{max}$ as well as $\Delta E_{min}$. The leading-order LHY correction to the chemical potential is $\Delta\mu=\frac{\partial\Delta E}{V\partial n}=\gamma_{QF}n^{3/2}$, where  $\Delta\mu_{max}=\frac{4}{3\pi^2}(\frac{M}{\hbar^2})^{\frac{3}{2}}[\Phi_{max}(k)]^{\frac{5}{2}}n^{\frac{3}{2}}$ and $\Delta\mu_{max}=\frac{4}{3\pi^2}(\frac{M}{\hbar^2})^{\frac{3}{2}}[\Phi_{min}(k)]^{\frac{5}{2}}n^{\frac{3}{2}}$, respectively. We extract the LHY coefficient $\gamma_{QF}$ from $\Delta\mu_{max}$ or $\Delta\mu_{min}$ as an approximated quantum fluctuation correction to the Rydberg-dressed system. Consequently, the dimensionless BGPE should be modified as:\par
\begin{center}
\begin{equation}
\begin{split}
i\frac{\partial\Psi(\bm{r},t)}{\partial t}\!=\!\left[\!-\frac{1}{2}\!\nabla^2\!+\!\Phi(\bm{r},t)\!+\!\gamma_{QF}\!|\Psi(\bm{r},t)|^3\!\right]\!\Psi\!(\bm{r},t),
\end{split}\tag{2}
\end{equation}
\end{center}
\noindent where $\Phi(\bm{r},t)={\alpha}|{\Psi}(\bm{r},{t})|^2-{\gamma}\int d^3{r}'\frac{|{\Psi}({\bm{r}},{t})|^2}{1-|{\bm{r}}-{\bm{r}}'|^6}$ with the contact interaction $\alpha=\frac{4\pi a_sN}{R_c}$, the long-range vdW interaction ${\gamma}=\frac{MN|\tilde{C}_6|}{\hbar^2R_c^4}$, and the quantum fluctuation part ${\gamma}_{QF}=\frac{4N^{\frac{3}{2}}}{3\pi^2\hbar^5}(\frac{M}{R_c})^{\frac{5}{2}}[\Phi_{max/min}(k)]^{\frac{5}{2}}$.
\par
In order to study the properties of individual droplets, we solve Eq. (2) in imaginary time for different interactions \cite{PhysRevA.93.061603,PhysRevA.88.033618,PhysRevA.102.063307}. The parameter $\eta=\gamma/\alpha$ is used to characterize the strength of the long-range attractive interaction with respect to the repulsive contact interaction. The corresponding energy functional can be expressed as $E=E_{MF}+E_{LHY}$, where $E_{MF}=\int[\frac{1}{2}{|\nabla\Psi(\bm{r})|^2}+\frac{1}{2}\alpha|\Psi(\bm{r})|^2]d\bm{r}-\frac{1}{2}\gamma\int[\frac{1}{1+|{\bm{r}}-{\bm{r}}'|^6}|\Psi^2(\bm{r})||\Psi^2(\bm{r}')|]d\bm{r}d\bm{r}'$ and $E_{LHY}=\int[\gamma_{QF}|\Psi(\bm{r})|^5]d\bm{r}$. Fig. 1(d) suggests that the LHY energy shares an opposite symbol from the MF energy, where the interaction strength $\eta$ and the particle number $N$ are changed, guaranteeing $E_{LHY}/E_{MF}$ negative. Meanwhile, the LHY energy and the MF energy are in the same order, so the LHY correction should not be neglected in the Rydberg-dressed case \cite{PhysRevLett.121.173403}.\par
\begin{figure}[t!]
\includegraphics[width=8cm]{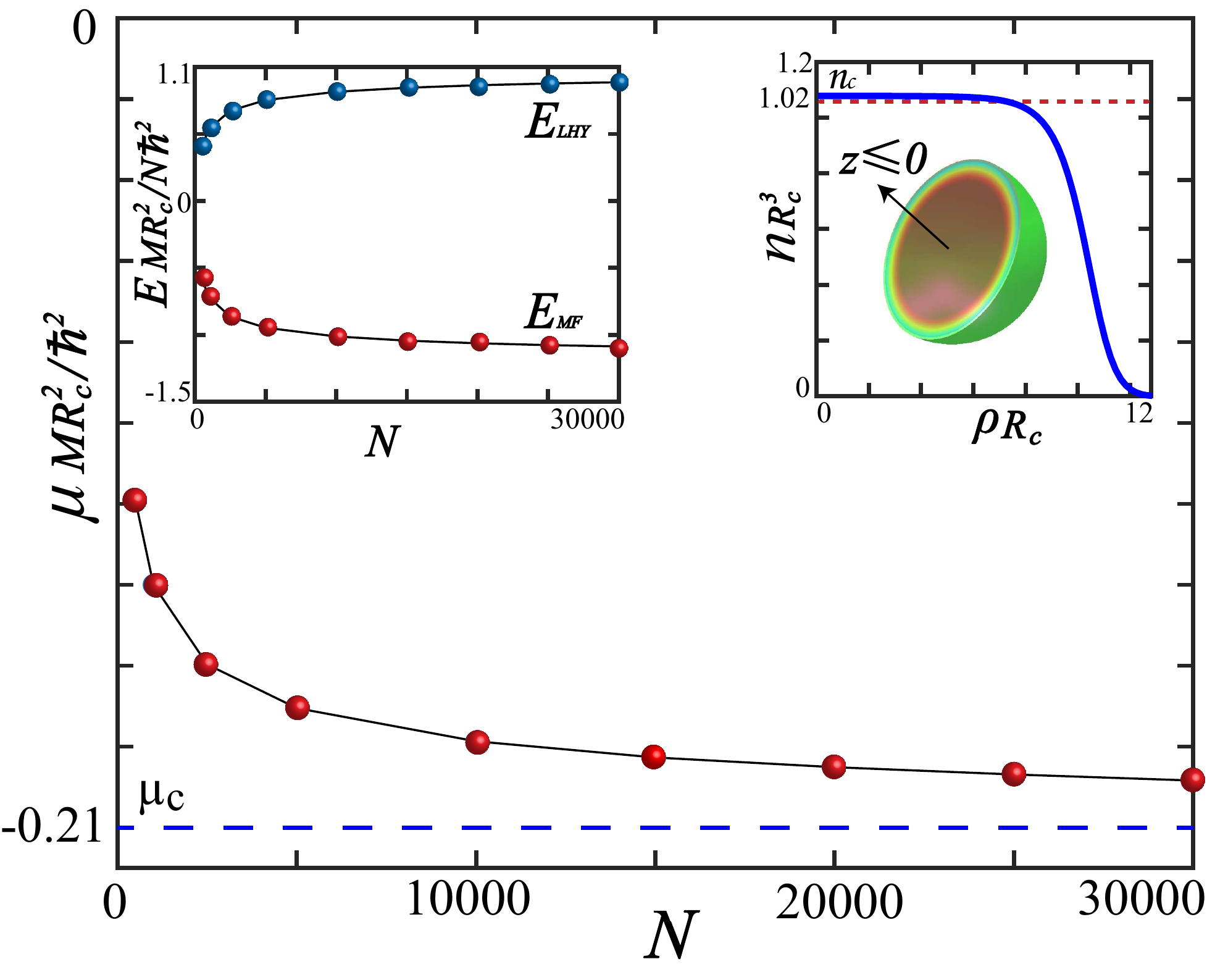}
\caption{Chemical potential $\mu$ of droplets as a function of the particle number $N$, where $\mu$ approximates the critical value $\mu_c=0.21$ (green dotted line). The energies of droplets are depicted in the left-pannel inset, where the LHY energy $E_{LHY}$ and the mean-field energy $E_{MF}$ are displayed from top to bottom. The density profile of a droplet with $N=10^4$ along the $\rho(x,y)$ direction and the corresponding density distribution for $z\le 0$ are shown in the right-pannel inset, where the peak value approximates the critical value $n_c=1.02$ (red dotted line).}
\label{fig2}
\end{figure}
\section{Droplets in the Rydberg-dressed system}
To verify the properties of the droplets, we compare the ground-state energy obtained from Eq. (2) with different number of atoms $N=\int |\Psi(\bm{r},t)|^2d^3r$. Compared with the Bose mixture droplets \cite{PhysRevLett.121.173403}, as shown in the left-pannel inset of Fig. 2, $E_{LHY}$ increases with $N$, while $E_{MF}$ is negative and decreases with $N$. This phenomenon can attribute to the vdW potential which replace the harmonic trap, providing attractive nonlocal nonlinearities that does not lead to condensate collapse \cite{Balewski_2014,PhysRevLett.106.170401,PhysRevA.76.054702}. The flat-top density distribution of droplets is shown in the right-pannel inset of Fig. 2, where the particle number $N=10^4$. Here the particle density {$n\sim 10^{15}/cm^3$}, the same order as the typical density in the 3D droplet experiment \cite{PhysRevLett.120.235301}. One can therefore apply Thomas-Fermi (TF) approximation to analyze them, i.e., neglecting the contribution from the kinetic term \cite{PhysRevA.98.063602}. Consequently, the energy of the system can be written as $E=\frac{1}{2}[-\epsilon n^2+\alpha n^2+\frac{4}{5}{\gamma_{QF}} n^{5/2}]V$, where $V=N/n$ denotes the volume of the droplet, $n=|\Psi(\bm{r},t)|^2$ is the ground-state density, $N$ is the total particle number, and $\epsilon=\gamma\int \frac{1}{1+|{\bm{r}}-{\bm{r}}'|^6}d\bm{r}d\bm{r}'$ is the long-range attraction strength. When the system is in equilibrium with the quantum pressure $\frac{dE}{dn}=0$, we have $\sqrt{n}=-\frac{5}{6{\gamma_{QF}}}(\alpha-\epsilon)$. Substituting $\epsilon\approx 6.6\gamma$ into the aforementioned density, we have the critical density peak $n_c=1.02$. Moreover, according to Eq. (2) we assume the solution $\Psi(\bm{r},t)=\phi(\bm{r}) e^{-i\mu t}$. With TF approximation, we have the chemical potential at equilibrium $\mu_c=(\alpha-\epsilon)n+{\gamma_{QF}} n^{3/2}=-0.21$ (green dotted line). Clearly, as shown in Fig. 2, the chemical potential $\mu$ of the droplets decreases with the total particle number $N$ approximating the critical $\mu_c$. The chemical potential is always negative, implying that the state is self-bound in the equilibrium. Note that the $\mu(N)$ curve satisfies the necessary stability condition in the form of the Vakhitov-Kolokolov criterion $d\mu/dN<0$ \cite{BERGE1998259}.
\par
\begin{figure}[t!]
\includegraphics[width=8.9cm]{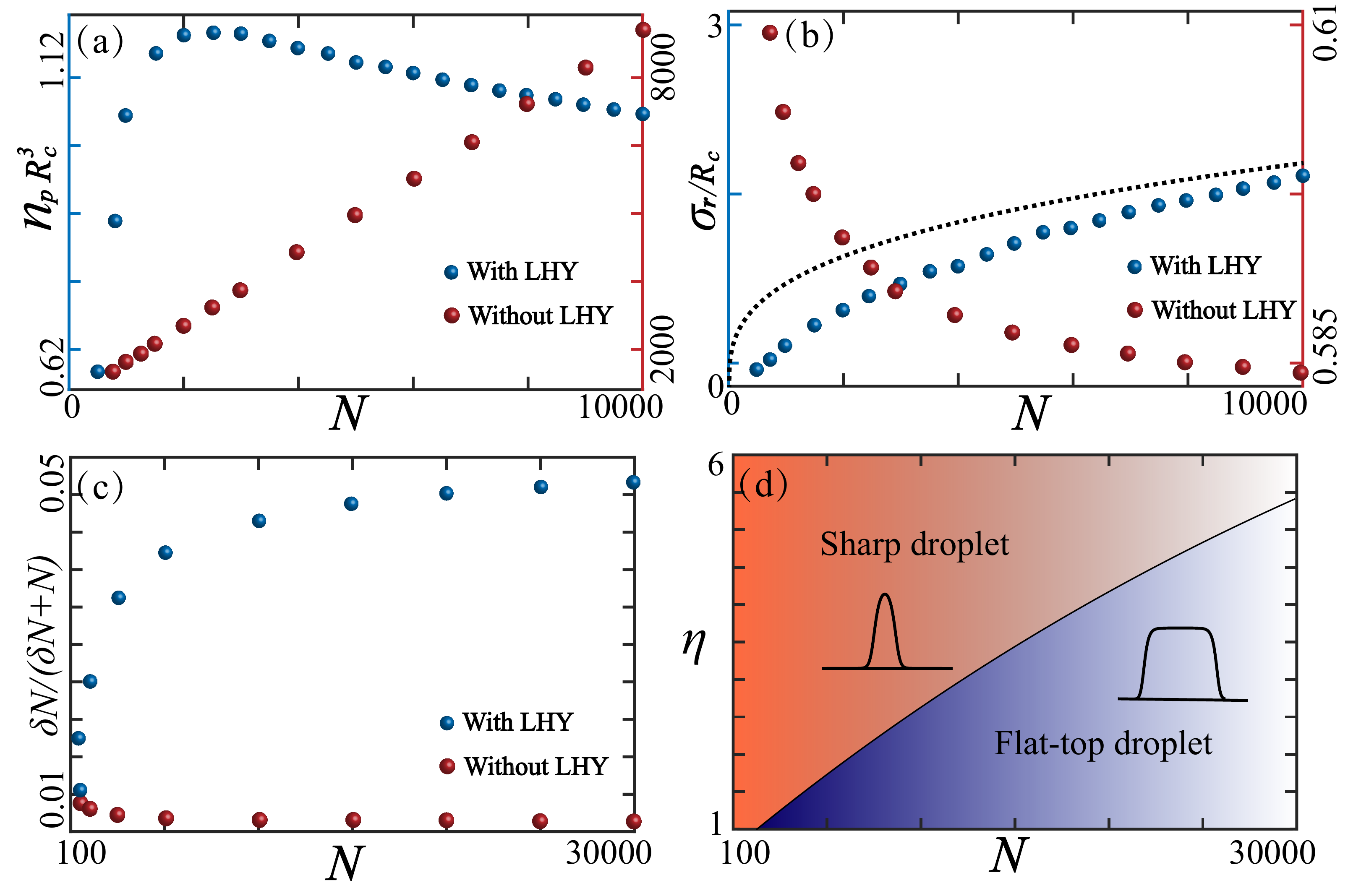}
\caption{(a) Peak density $n_p$ of droplets (with LHY correction) and bright solitons (without LHY correction) share different behaviors related to the evolution of the particle number $N$ (the interaction ratio $\eta=2$). (b) Width $\sigma_r$ of droplets and bright solitons share different behaviors with related to the evolution of the $N$. (c) Quantum depletion $\delta N/(\delta N+N)$ as a function of $N$ (with/without LHY correction). (d) Phase diagram of the self-trapping droplets where sharp droplets and flat-top droplets are distinguished by $\eta$ and $N$.}
\label{fig3}
\end{figure}
The evolution of droplets with different particle numbers exhibits completely different behaviors from bright solitons. Fig. 3(a) depicts the condensate peak density $n_p$ as a function of particle number $N$ where the interaction ratio $\eta=2$. Neglecting the LHY correction, the self-trapped bright soliton is balanced by the attractive vdW interaction and the repulsive contact interaction, that causes $n_p$ increasing linearly with $N$. However, when LHY correction is under consideration, $n_p$ increases with $N$ for sufficiently low particle numbers, which exhibits self-trapped sharp droplets \cite{PhysRevA.98.013612}. The peak density $n_p$ drops slightly towards the critical density $n_c$ for sufficiently larger $N$, and the density exhibits flat-top shape. The corresponding width of the self-trapped condensate is exhibited in Fig. 3(b) where the appearance of LHY correction is also discussed. The widths of the condensate can be extracted from the BGPE solutions employing $\sigma_r^2=c\int \bm{r}|\Psi(\bm{r})|^2d\bm{r}$, where $c$ is normalization constant. Without the LHY correction, the width of the bright soliton decreases with $N$. On the contrary, when considering LHY correction, the width of the droplet increases smoothly with $N$. {Through substituting the particle density into the energy functional $E$ and solve the equation $dE/dV=0$, we can obtain the relationship between the volume and particle number $V=\frac{36}{25}\gamma_{QF}^2\frac{N}{(\alpha-\epsilon)^2}$. Assuming the droplet is a 3D isotropic ball, we can easily find the relationship between the width and the total particle number of the droplet $\sigma_r=\left[\frac{27{\gamma_{QF}}^2 N}{25\pi(\alpha-\epsilon)^2}\right]^{\frac{1}{3}}$ (exhibiting by the green line). There is a obvious gap between the numerical and the analysis results for small $N$ where the sharp droplet prevails among the ground states. However, the gaps become small for large $N$ where the flat-top droplet dominates the ground states.
\par
The validity of the beyond mean-field approximation can be checked by evaluating the quantum depletion \cite{PhysRevA.98.033612}. According to the Bogoliubov theory, the fluctuation part $\delta{\Psi}(\bm{r},t)$ around the condensate can be subjected to a canonical transformation resulting in the expansion $\delta{\Psi}(\bm{r},t)=\sum_k u_k(\bm{r})e^{-i\omega_k t}+v_k^{\ast}(\bm{r})e^{i\omega_k t}$ \cite{PhysRevA.87.061602,PhysRevB.86.060510,PhysRevResearch.2.033522}. The mode functions $u_k(\bm{r})$, $v_k(\bm{r})$ and the collective frequencies $\omega_k$ are determined by the Bogoliubov-de Gennes (BdG) equation. At zero temperature, the number of the non-condensate particles can be calculated by $\delta N=\int |u_k|^2+|v_k|^2d\bm{r}$, where $k$ is restricted by the nonnegative mode frequencies $\omega_k>0$. It is convenient to numerically solve the BdG equation in the Fourier space. As shown in Fig. 3(c), quantum depletion $\delta N/(\delta N+N)$ decreases with the total particle number $N$ for bright solitons while increases with $N$ for droplets. One can see that, the quantum depletion is always less than $5\%$, thereby confirming the validity of the MF/BMF approximations. We further summarize the phase diagram of droplets in the Rydberg-dressed condensates in Fig. 3(d) with the total particle number $N$ and the interaction ratio $\eta$. Through making a fully numerical calculation for $\eta\in[1,6]$ and $N$ up to $3\times 10^4$, we find that the sharp droplet appears for small $N$ while the flat-top droplet appears for large $N$. For a fixed $\eta$, the peak density of sharp droplets increases with the particle number booming. On the contrary, the density peak of flat-top droplets decreases with particle number booming.
\par
\section{Dynamical Stabilities and Collisions of Droplets}
\begin{figure}[t!]
\includegraphics[width=8.6cm]{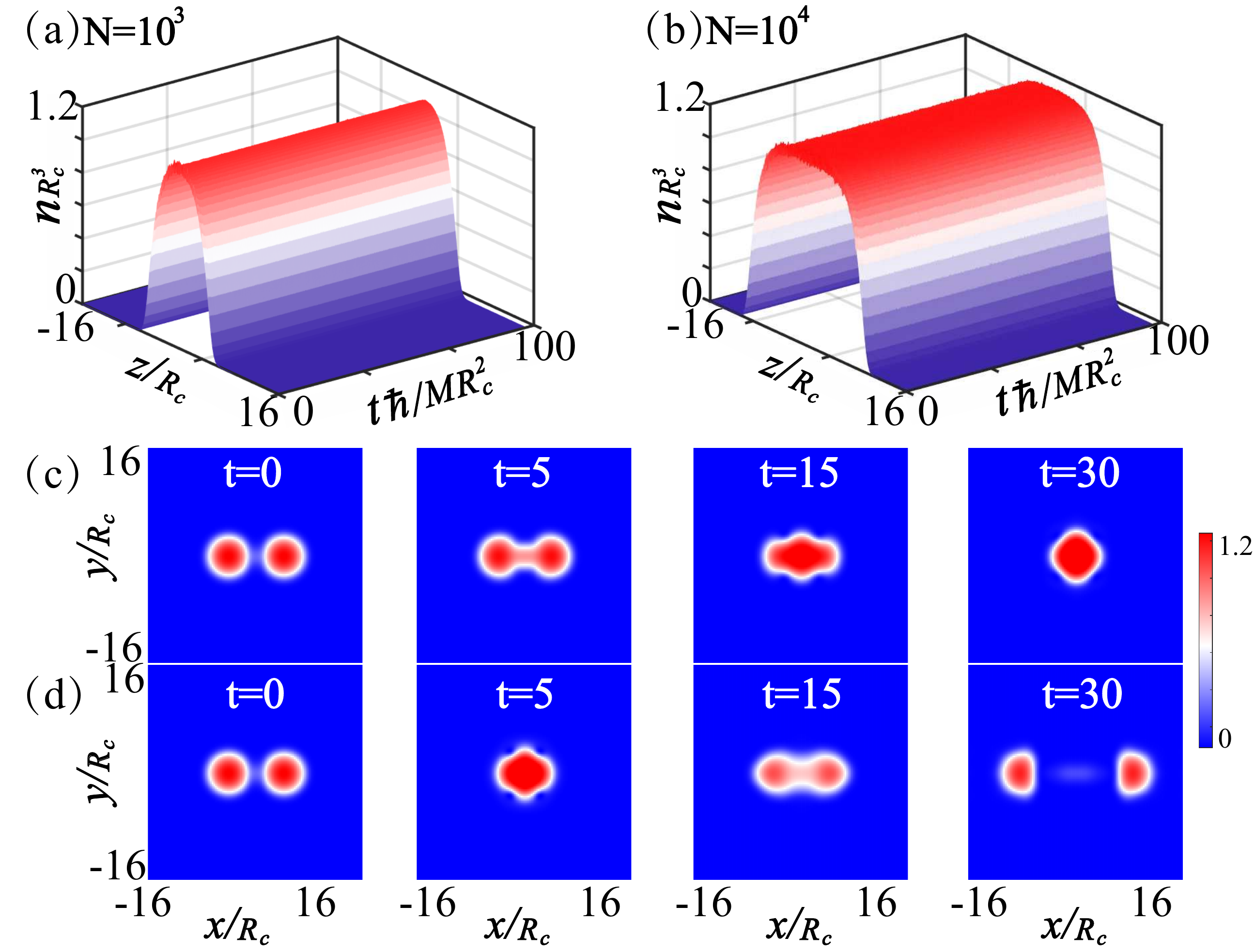}
\caption{(a-b) Real-time dynamic evolution of the sharp droplet and the flat-top droplet solution in the presence of Gaussian noise at $3\%$ level. Densities $n(0,0,,z)$ of the droplet vary with time, which indicates the long-time stability. The particle numbers are $N=10^3, 10^4$, respectively. (c-d) Inelastic and quasielastic collides of 3D droplets with different initial momentum $k=0.1$, $0.5R_c^{-1}$, respectively.}
\label{fig4}
\end{figure}
Experimentally, coherent Rydberg excitation of cold atoms has been demonstrated under various conditions \cite{Balewski_2014}. In order to demonstrate the dynamic
stability of the droplets, Figs. 4 (a-b) depict the real time evolution of the droplet solutions obtained through imaginary-time propagation method. After adding $3\%$ Gaussian noise at time $t=0$, the droplets are released freely. From the long-time evolution of the droplets, the density $n(0,0,z)-t$ images indicate that both the
sharp droplet ($N=10^3$) and the flat-top droplet ($N=10^4$) are stable. Finally, to consider the collisions between moving droplets, we place two droplets centered at initial positions $(x,y,z)=(\pm x_0,0,0)$ without harmonic trap. The droplets then start moving to collide at the center with the same initial condition except the center-of-mass kinetic energy $k=0.1$, $0.5R_c^{-1}$, respectively. In Fig. 4(c), the slowly moving droplets feature a inelastic collision. However, for larger initial kinetic energies in Fig. 4(d), the collision approaches to quasielastic.
\par
\section{Conclusion}
We firstly propose a Rydberg-dressed system as a platform for the realization of isotropic three-dimensional droplets. The isotropy of droplets may be broken when considering $nP$ or $nD$ states Rydberg-dressing \cite{PhysRevA.105.013326,PhysRevB.106.024506,PhysRevA.104.L061302}. The quantum fluctuation effect, i.e., LHY correction, which turns bright solitons into droplets, should not be neglected in the Rydberg-dressing system. We summarize the phase diagram of droplets with various particle numbers and interaction ratios. Finally, we exam the dynamic stability and collision of droplets which exhibits long lifetime stability. Considering the potential applications of droplets to investigate superfluid or polaron physics \cite{PhysRevB.102.060502,Ardila_2019}, it worth further introducing spin-orbit coupling \cite{PhysRevLett.117.185301,PhysRevA.104.053325} or spin-angular momentum \cite{PhysRevLett.122.110402} into the Rydberg-dressed droplets.
\par
This work was supported by National Key $\rm{R\&D}$ Program of China under grants $\rm{Nos.\,2021YFA1400900}$, $\rm{2021YFA0718300}$, $\rm{2021YFA1402100}$, NSFC under grants $\rm{Nos.\, 61835013}$, $\rm{12234012}$, $\rm{11874064}$, $\rm{11934014}$, $\rm{12004392}$, Space Application System of China Manned Space Program, Strategic Priority Research Program of the Chinese Academy of Science (CAS) under grant $\rm{No.\, XDB21030300}$, and NKRD Program of China under grant $\rm{No.\, 2021YFA1402104}$.
\bibliography{mainHu}
\end{document}